\documentstyle[psfig]{caosp}
\begin{document}

\htitle{1996 multisite photometric and spectral ...}
\hauthor{D.E. Mkrtichian {\it et al.}}

\title { 1996 multisite photometric  and spectral campaign on the
$\lambda$\,Boo star 29 Cyg}

\author{D.E.~Mkrtichian \inst{1} \and A.V.~Kusakin \inst{2} \and
V.P.~Malanushenko \inst{3} \and M.~Papar\'o \inst{4} \and M.C.~Akan
\inst{5} \and J.R.~Percy \inst{6} \and S.~Thompson \inst{6} \and
K.~Krisciunas \inst{7} \and V.I.~Burnashev \inst{3} \and C.~Ibanoglu
\inst{5} \and R.~Pekunlu \inst{5} \and A.~Devlen \inst{5}  A.~Ozturk
\inst{5}  \and V.A.~Koval \inst{2}}
\institute{Astronomical Observatory, Odessa State University,
Shevchenko Park, Odessa, 270014, Ukraine \and Sternberg State Astronomical
Institute, Universitetsky prospect, 13, Moscow, 119899, Russia
\and Crimean Astrophys. Observatory, Nauchny, Crimea, Ukraine, 334413
\and Konkoly Observatory, P.O. Box 67,H-1525, Budapest, Hungary
\and Ege University Observatory, Bornova, Izmir, Turkey
\and Erindale College, University of Toronto, Mississauga, Ontario,
Canada L5L 1C6
\and Now at Astronomy Department, University of Washington, Box 351580,
Seattle, Washington 98195 USA}

\maketitle

\section{Introduction}
\label{intr}
\vspace{-2mm}
29\,Cyg is the first among $\lambda$\,Boo-type stars for which
definite evidence of pulsations has been found (Gies\,\&\,Percy, 1977).
Since the discovery of variabilty, 29\,Cyg has been studied photometrically by
Cooper \& Walker (1994), Kusakin\\
\& Mkrtichian (1996), Paunzen\,\&\,Handler (1996).

The 1996 multisite two-stage  photometric and spectral campaign on
29 Cyg was organized by the CAN group (see Mkrtichian et al., 1998)
between 5--17 August and 15--25 September against the background of
continuous monitoring of the star from the CAN site TSAO between 27 July and
30 September. This campaign was a natural resumption of the 1995 photometric
investigation of 29 Cyg (Kusakin\,\&\,Mkrtichian, 1996).

\vspace{-2mm}
\section{Photometry}
\vspace{-1mm}
 A total of 48 photometric and 2 spectroscopic
nights at observatories in Ukraine, Kazakhstan, Turkey, Hungary, Canada and
USA were acquired during the campaign.
29 Cyg is known as a star with variable amplitudes of modes
(Kusakin \& Mkrtichian, 1996). In this note, we focus on the search for
a solution with a stationary frequency for the interval
JD\,2450292 - JD\,2450357.
The steps of the  preliminary DFT analysis of combined  V-filter
data of 45 selected nights, made using the pre-whitening technique, are shown
in Figure 1; the frequencies found are given in Table 1.

\begin{table}
\small
\begin{center}
\caption{Frequency solution for JD 2450292 - 2450357 V-data set}
\label{fre}
\begin{tabular}{|ccc|}
\hline
&Frequency& Semi-ampl. (V) \\
&$d^{-1}$& mag \\
\hline \it $f_{1}$&
37.425&0.0102\\
\it $f_{2}$& 34.721&0.0038\\
\it $f_{3}$&29.774&0.0030\\
\it $f_{4}$& 28.160&0.0023\\
\it $f_{5}$&25.189&0.0021\\
\it $f_{6}$& 33.629&0.0023\\
\hline \end{tabular}
\end{center}
\end{table}

\begin{figure}[hbt]
\centerline{\psfig{figure=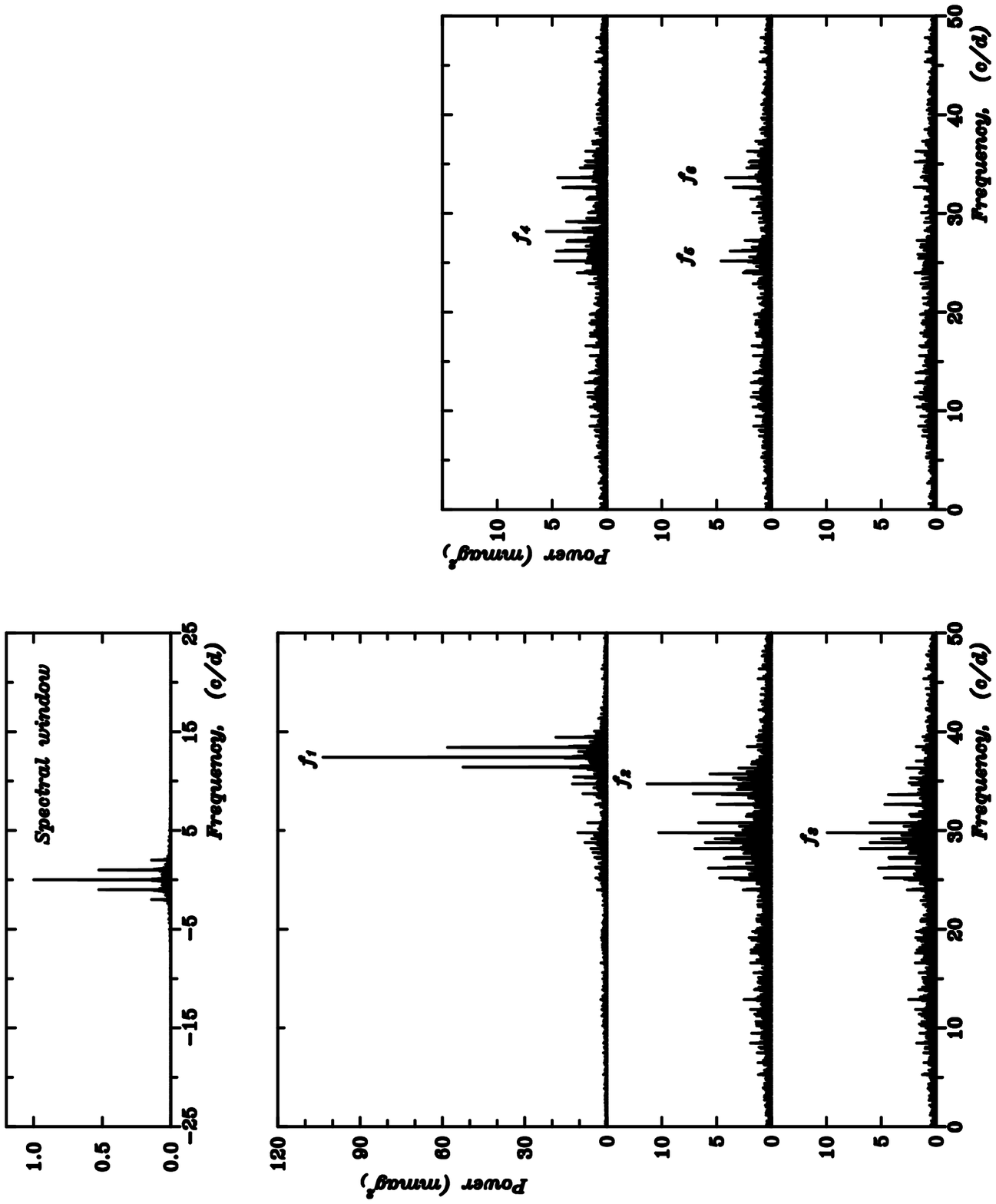,width=5cm,angle=270}}
\caption{}
\end{figure}

Comparison with the yet unpublished analysis of our 1995 V data shows
conspicious peculiarities:
\begin{itemize}
\item  In 1996 the main mode of pulsation, with the frequency 37.425~d$^{-1}$
($\Delta$V=0.020), did not change its amplitude, in comparison with the 1995
V-band observation ($\Delta V=0.020$)  obtained at TSAO.
\item The second main mode, with $\nu=29.77$~d$^{-1}$, decreases its amplitude
from $\Delta V= 0.012$ to $\Delta V=0.006$, i.e. by a factor of two.
\item Evidence of additional low-amplitude modes in the power
spectrum can be suspected in the range $10-40$~d$^{-1}$.
\end{itemize}

\vspace{-4mm}
\section{Spectroscopy:}
\vspace{-2mm}
When our work on 29 Cyg started, only one $\lambda$\,Boo-type star, HD 111604,
was investigated spectroscopically for pulsations (Bohlender et al., 1996).
The new spectral observations of 29\,Cyg were carried out at the coud\'{e}
focus of the 2.6-m telescope of the Crimean  Astrophysical Observatory
during two consecutive nights, August 11/12 and 12/13 1996.
We present in Figure 2a,b the radial
\begin{figure}[hbtp]
\parbox[b]{12cm}{\psfig{figure=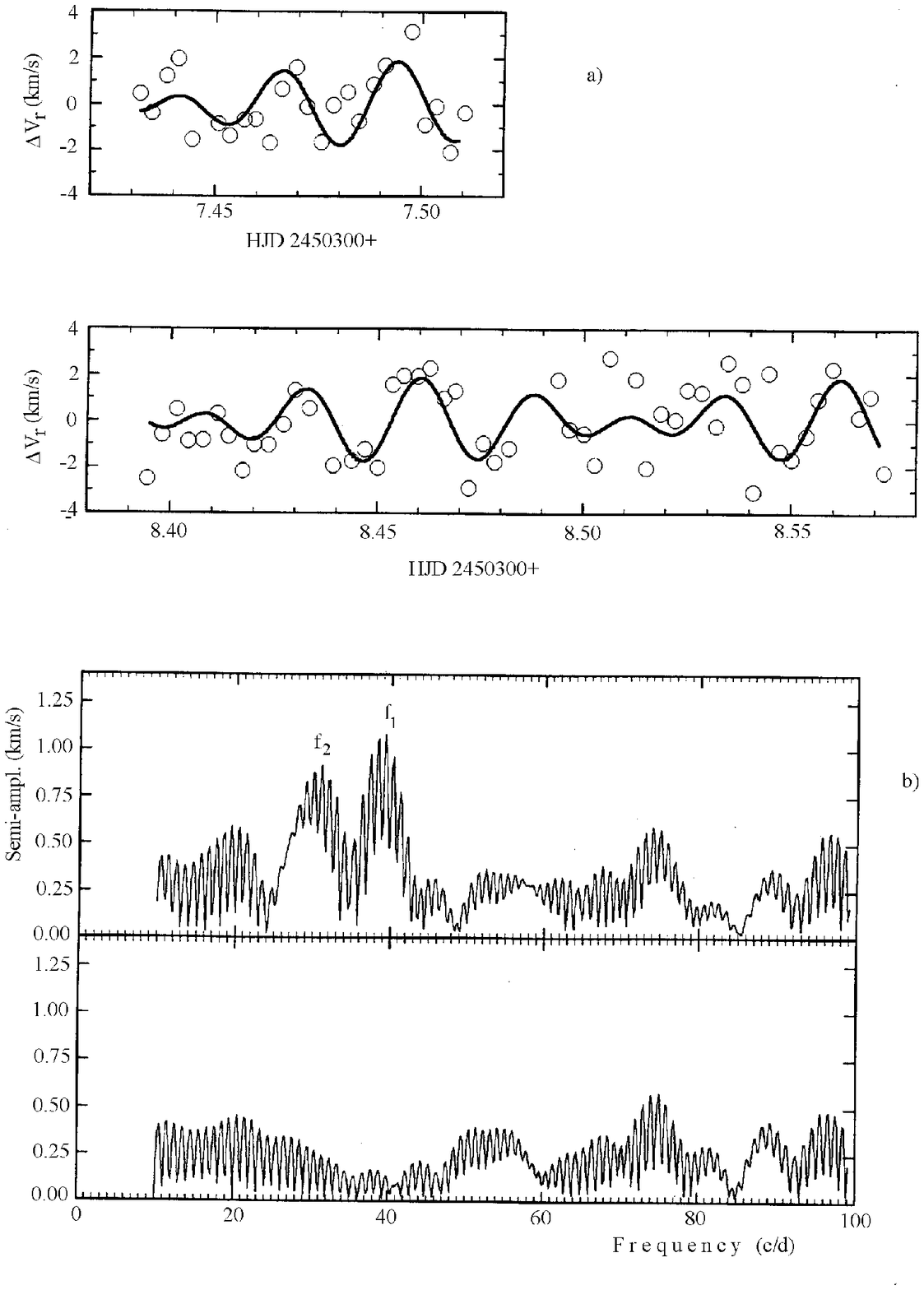,width=6cm,angle=1}}
\parbox[b]{6cm}
{velocity (RV) variations of
the H$_{\alpha}$ line (the dots) for two nights, JD\,2450307 and JD\,2450308
and two steps of DFT frequency analysis in combination with prewhitening
technique.
 RV variations with peak-to-peak amplitudes of about 4 km\,s$^{-1}$
are clearly seen in the figure. The main peak at 39.3~d$^{-1}$ with a
semi-amplitude of 1.0 km\,s$^{-1}$ in amplitude spectra, corresponds, within
the uncertainties, to the 37.425~d$^{-1}$ frequency of photometric variability
having the highest amplitude.
The velocity-to-amplitude ratio for 37.425~d$^{-1}$ is
$2K/\Delta V = 100$~km\,s$^{-1}$\,mag$^{-1}$.
The second frequency $f_{2}=30.0$~d$^{-1}$ with a semi-amplitude
of 0.8 km\,s$^{-1}$ is more probably an unresolved combination of the two
known photometric frequencies, 34.72~d$^{-1}$ and 29.77~d$^{-1}$.}
\small
\makebox[2.4cm][l]\, {\bf Figure 2.}
\end{figure}

\vspace{-4mm}
\section{Further work and prospect:}
\vspace{-2mm}
We plan to do analysis of the simultaneous multicolor W,B,V,R
 ~TSAO observations for the mode-identification of the highest-amplitude
modes. A joint fine analysis of photometry and spectroscopy is planned.
In July-October 1997 the second multisite photometric and
simultaneous spectral campaign on 29 Cyg was undertaken by CAN and
the observatories in Arizona and Turkey. Data were acquired during more than
82 photometric and 2 spectroscopic nights. The combined analysis of
1995/1996/1997 data
will allow the investigation of the amplitude variability of modes over
3 years, a reduction of the noise level in the power spectrum and the
extraction of small amplitude modes suspected in the 10-40 d$^{-1}$ range.

\vspace{-3mm}
\acknowledgements
The spectral observations described in this publication were made possible
in part thanks to Grants R2Q000 and U1C000 from the International Science
Foundation and to Grant A-05-067 from the ESO \& CEE Programme.


\vspace{-4mm}

\begin{thebibliography}{}

\article{Bohlender, D.A., Gonzalez, J.-F., Kennelly, E.J.}{1996}{\aaa}{307}{L9}
\bibitem{} Cooper,\,W.A., Walker,\,E.N.:\, 1994, {\it Getting the Measure of
the Stars}, Mir,\,Moscow,\,271
\article{Gies, D.R., Percy, J.R.}{1977}{\aj}{82}{166}
\bibitem{} Kusakin, A.V., Mkrtichian, D.E.: 1996, {\it IBVS}, 4314
\bibitem{} Mkrtichian, D.E., Kusakin A.V., Janiashvili E.B., Lominadze J.G.,
 Kuratov K., Kornilov V.G., Dorokhov N.I., Mukhamednazarov S.: 1998,
{\it these Proceedings}, 238
\bibitem{} Paunzen, E., Handler, H.: 1996, {\it IBVS}, 4318

\end{thebibliography}
\end{document}